# First Observation of a Type II Solar Radio Burst Transitioning between a Stationary and Drifting State

Nicolina Chrysaphi[1], Hamish A. S. Reid[1,2], and Eduard P. Kontar[1]
[1] School of Physics & Astronomy, University of Glasgow, Glasgow, G12 8QQ, UK; n.chrysaphi.1@research.gla.ac.uk
[2] Department of Space and Climate Physics, University College London, London, RH5 6NT, UK

## Abstract

Standing shocks are believed to be responsible for stationary Type II solar radio bursts, whereas drifting Type II bursts are excited by moving shocks often related to coronal mass ejections (CMEs). Observations of either stationary or drifting Type II bursts are common, but a transition between the two states has not yet been reported. Here, we present a Type II burst which shows a clear, continuous transition from a stationary to a drifting state, the first observation of its kind. Moreover, band splitting is observed in the stationary parts of the burst, as well as intriguing negative and positive frequency-drift fine structures within the stationary emissions. The relation of the radio emissions to an observed jet and a narrow CME were investigated across multiple wavelengths, and the mechanisms leading to the transitioning Type II burst were determined. We find that a jet eruption generates a streamer-puff CME and that the interplay between the CME-driven shock and the streamer is likely to be responsible for the observed radio emissions.

*Unified Astronomy Thesaurus concepts:* Solar physics (1476); Solar activity (1475); Solar radiation (1521); Solar coronal mass ejection shocks (1997); Shocks (2086); Radio bursts (1339); Solar coronal radio emission (1993)

## 1. Introduction

Type II solar radio bursts have long been associated with shock wave formations in the heliosphere and have thus been frequently linked to solar eruptive events like coronal mass ejections (CMEs) (see, e.g., Pikel'Ner & Gintsburg 1964; Dulk et al. 1971; Leblanc et al. 2000; Cho et al. 2007; Kouloumvakos et al. 2014; Zucca et al. 2014; Chrysaphi et al. 2018; Maguire et al. 2020). Shocks preceding CMEs propagate through the heliosphere exciting Langmuir waves which, through the plasma emission mechanism, produce radio radiation that manifests in dynamic spectra as slowly drifting structures with characteristic frequency-drift rates of $\lesssim -1$ MHz s$^{-1}$ (Wild & McCready 1950; Roberts 1959; McLean & Labrum 1985). These radio emissions are referred to as drifting Type II bursts and are thought to reflect the speed with which their exciter, i.e., the moving shock, propagates in the corona and in the interplanetary medium (McLean & Labrum 1985). Although not always the case, Type II emission sources are believed to be at the flanks of their associated CMEs (see e.g., Cho et al. 2007; Carley et al. 2013; Zucca et al. 2014; Chrysaphi et al. 2018; Krupar et al. 2019; Mancuso et al. 2019; Morosan et al. 2019).

The location(s) and way that radio sources form on a shock wave are debated and current interpretations are disputed. A puzzling characteristic of Type II bursts known as "band splitting" is used to challenge the proposed models for the excitation of radio emission on shocks (Roberts 1959; McLean 1967; Smerd et al. 1974, 1975; Holman & Pesses 1983; McLean & Labrum 1985; Mann et al. 1995, 2018; Vršnak et al. 2001; Zimovets et al. 2012; Du et al. 2015; Chrysaphi et al. 2018). Band splitting refers to the splitting of any of the harmonic bands of a Type II burst into (usually two) thinner lanes, or "subbands," but the current understanding of physical mechanisms fails to fully and accurately describe this morphological characteristic. While the ability to image radio emissions with a higher resolution has improved over the past decade, it is still difficult to distinguish between the most widely accepted band-splitting interpretations that disagree on the origin of the subband emission sources with respect to the shock front (see, e.g., Smerd et al. 1974, 1975; Holman & Pesses 1983). This is mainly because observations of radio emissions do not represent the intrinsic properties of the radio sources since radio-wave propagation effects, the dominant of which is believed to be scattering (Kontar et al. 2017), alter the intrinsic properties necessary for understanding the physical mechanisms exciting the emissions (Kontar et al. 2019). Like all plasma or harmonic radio emissions, the observed positions of Type II bursts are affected by scattering effects (Smerd et al. 1975), but a quantitative estimation of their extent had not been conducted until recently. Chrysaphi et al. (2018) derived a simple, analytical expression to quantitatively estimate the radial shift caused by scattering. By accounting for this effect in a split-band Type II burst, they found that significant separations observed between higher- and lower-frequency subband sources are negated, dramatically altering the results with respect to what would be concluded if scattering effects were to be ignored. Consequently, they highlighted the importance of accounting for radio-wave scattering effects when interpreting observations of Type II bursts.

While the majority of Type II bursts recorded are drifting bursts, and thus related to moving shocks, Type II bursts that show little or no drift on average have also been reported (see Aurass et al. 2002; Aurass & Mann 2004; Mel'nik et al. 2004; Mann et al. 2009; Chen et al. 2019). These are known as "stationary" (or quasi-stationary) Type II bursts (Aurass et al. 2002) and indicate that the emitting source has not migrated to a location in the heliosphere with a different local plasma density. Such emissions are often interpreted as signatures of standing shocks related to solar flares, known as termination shocks (Aurass et al. 2002; Aurass & Mann 2004; Mann et al. 2009; Chen et al. 2019).

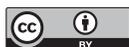

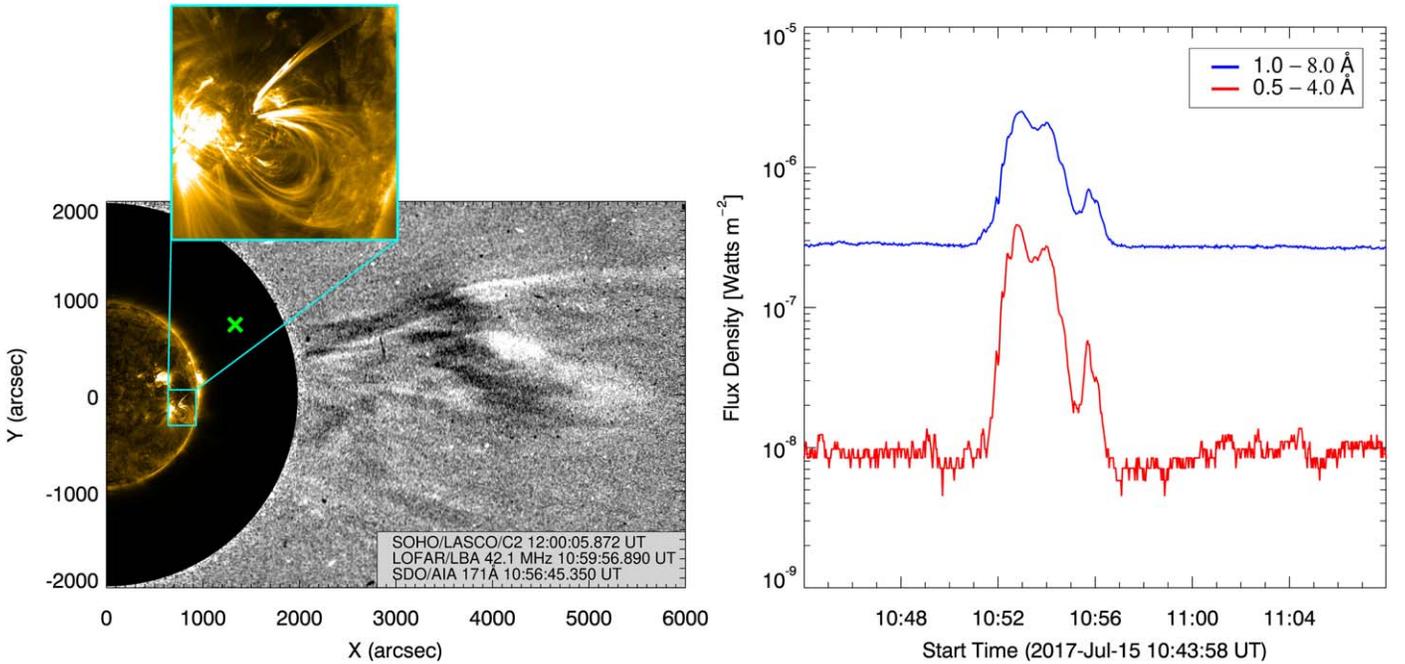

**Figure 1.** The left panel is a combination of LOFAR data, SOHO/LASCO/C2 running-difference data, and SDO/AIA (171 Å) data. The green cross illustrates the location of the Type II burst emissions and the black disk represents the occulting disk of the C2 coronagraph. The two CME fronts can also be distinguished, with brighter structures in LASCO's field of view (FOV) reflecting relative increases in intensity and darker structures reflecting relative decreases in intensity. The inset shows the coronal jet emerging from the northern edge of the active region. The right panel shows the true X-ray flux density measured by the GOES-15 XRS instrument during the jet's eruption at 0.5–4.0 Å (red curve) and 1.0–8.0 Å (blue curve).

Here, we present the first observation of a Type II solar radio burst that transitions between these two distinct states, specifically, from a stationary state to a drifting state, raising new questions as to the way that stationary Type II bursts can be formed. The presented Type II burst also undergoes band splitting during its stationary part. In combination with the transitioning state of the burst, the morphological characteristics of the presented event could improve our understanding of how and where Type II emissions are created with respect to the shock. As a first step, this study combines multiwavelength data to explore the mechanisms related to the Type II emissions observed, as well as the way in which a Type II burst with such morphology can form. An overview of the multiwavelength observations and how they were conducted is provided in Section 2. Section 3.1 presents the spectroscopic observation of the transitioning Type II burst and an analysis of its morphological features. Complementary observations to the radio emissions are described in Section 3.2. We focus on white-light coronagraphic images that depict the CMEs related to the radio emissions, as well as on the jet that is believed to have triggered the CMEs. A detailed description and analysis of the properties of the jet are provided. The spatial behavior of the radio sources before, during, and after the transition from a stationary to a drifting state is examined in Section 3.3. A presentation of intriguing fine structures within the stationary part of the Type II burst is also included, as well as imaging of a Type III burst occurring during the stationary Type II emissions. A summary of the observations and the interpretation of the mechanism generating the characteristic radio emissions is presented in Section 4.

## 2. Overview of the Observations

A Type II solar radio burst that experiences a transition from a stationary to a drifting state, as well as band splitting, was observed by the LOw-Frequency ARray (LOFAR; van Haarlem et al. 2013) between ∼30 and 70 MHz at ∼11:02 UT on 2017 July 15. To image the radio emissions, LOFAR's outer Low-Band Antenna (LBA) core stations were used in tied-array beam mode (van Haarlem et al. 2013), forming 217 individual beams which created a mosaic covering the Sun up to ∼2.7 $R_\odot$. For the calibration of the flux, Tau A observations were used before and after the solar observation, similar to Kontar et al. (2017) and Chrysaphi et al. (2018). The observation was conducted using 24 core stations providing a temporal resolution of ∼0.01 s, a spectral resolution of ∼12 kHz, and a synthesized beam with an FWHM of ∼10' at 30 MHz (for details see the Methods in Kontar et al. 2017). The sensitivity of such a configuration is ≲0.03 sfu per beam. For the analysis and presentation of the radio emissions, the spectral and temporal resolutions were rebinned and decreased to ∼73.2 kHz and ∼0.21 s, respectively.

In close temporal and spatial proximity to the Type II emissions, a coronal jet eruption was observed on the edge of an active region west of the solar limb at ∼10:51 UT in EUV images obtained from the Solar Dynamics Observatory (SDO; Pesnell et al. 2012) Atmospheric Imaging Assembly (AIA; Lemen et al. 2012), which observes with a ∼12 s cadence. The eruption of the jet was also observed in X-ray data taken by the Geostationary Operational Environmental Satellite (GOES-15) X-Ray Sensor (XRS) (Thomas et al. 1985; Garcia 1994), as indicated bin Figure 1.

Following the eruption of the jet, ejecta appeared at ∼11:12 UT in white-light coronagraphic images gathered by the Large Angle Spectroscopic Coronagraph (LASCO; Brueckner et al.

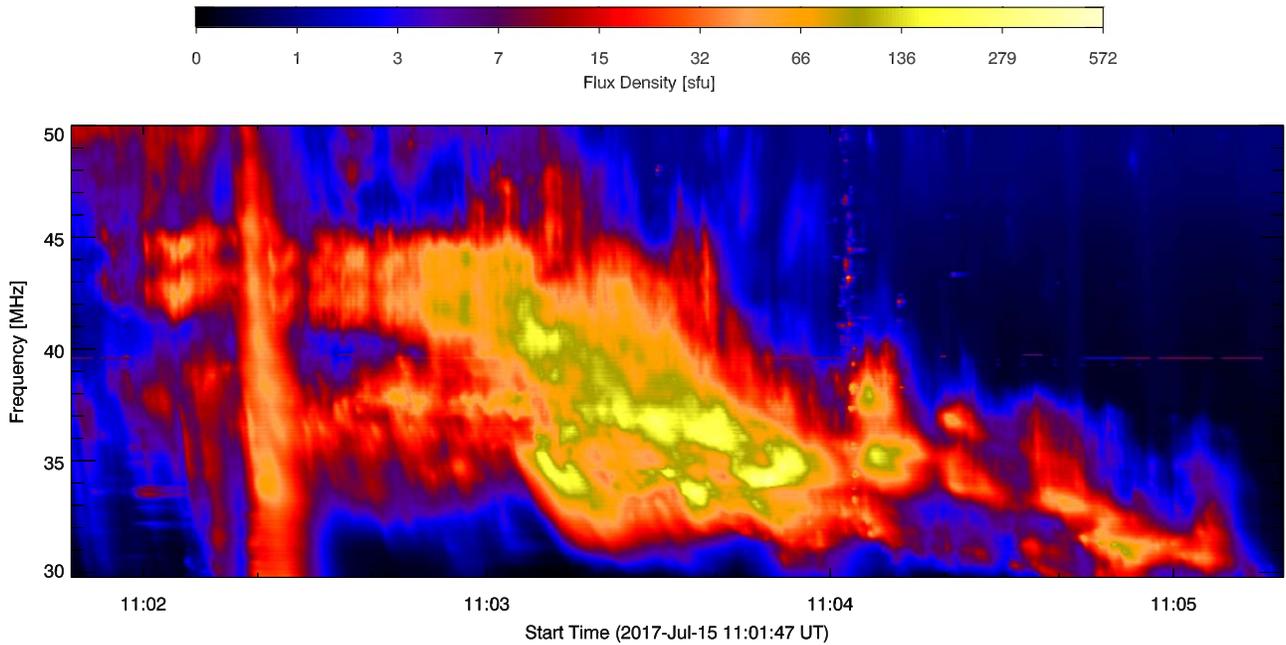

**Figure 2.** Dynamic spectrum depicting a Type II solar radio burst transitioning between a stationary and drifting state, as observed on 2017 July 15 by LOFAR's LBA stations. The stationary Type II part is observed between ∼11:02 and 11:03 UT and the drifting part between ∼11:03 and 11:05 UT. A Type III burst is also observed (at ∼11:02:20 UT) during the stationary Type II emissions. Prior to plotting, the spectral and temporal resolutions of the data were decreased from ∼12.2 kHz and ∼0.01 s to ∼73.2 kHz and ∼0.21 s, respectively.

1995) C2 camera on board the Solar and Heliospheric Observatory (SOHO; Domingo et al. 1995), which images heights between ∼2.2 and 6 $R_\odot$ with a ∼12 minutes cadence. The ejecta were identified as two CMEs that were temporally and spatially related to the observed radio emissions (see Figure 1).

### 3. Analysis of the Observations

#### 3.1. Spectroscopic Radio Observations

The observed Type II burst, shown in Figure 2, transitions between a stationary and a drifting state, a behavior never reported before. The stationary Type II emissions consist of two bands that appear at the same time, each experiencing band splitting. The relation of these two pairs of subbands—seen in Figure 2 at ∼35–39 MHz and ∼41–45 MHz, respectively—cannot be explained by harmonic plasma emission, given that a 1:2 frequency ratio is not observed (McLean & Labrum 1985; Mann et al. 1995). Each pair of subbands is either the result of an individual shock, or the Type II burst experiences simultaneous band splitting at two different locations, i.e., both pairs of subbands are the result of a single shock front. Should the latter be true, it would add another intriguing aspect to this observation, since simultaneous band-splitting structures in a single Type II burst are not believed to have been previously recorded. We note that the similarity in the temporal and morphological appearance of the emission patterns between the two pairs of subbands indicates that the regions of the shock (or shocks) exciting radio emissions simultaneously propagate through the same density region in the corona (Vršnak et al. 2001).

The first pair of subbands appears around 44 and 42 MHz (for the higher- and lower-frequency component, respectively), with an estimated average frequency split $\Delta f/f \approx 0.05$. The second pair of subbands appears around 37.5 and 36 MHz, with $\Delta f/f \approx 0.04$. These frequency-split values are somewhat lower than the typical range for drifting Type II bursts that experience band splitting ($\Delta f/f = 0.1$–0.5), which is in itself a puzzling characteristic of Type II bursts (Vršnak et al. 2001; Du et al. 2015). The drifting part of the Type II burst experiences a frequency drift at the rate of ∼−0.14 MHz s$^{-1}$, which is typical of Type II bursts (McLean & Labrum 1985; Mann et al. 1995). A Type III burst with a frequency-drift rate of ∼−5 MHz s$^{-1}$ is is observed at ∼11:02:20 UT—during the stationary Type II emissions.

#### 3.2. EUV and White-light Observations

The Type II emissions are temporally and spatially associated with a jet eruption and coronal ejecta identified as two CME fronts. A solar flare of magnitude C1.4 occurred at ∼10:50 UT and was preceded by the ejection of the jet which lasted in the AIA 171 Å channel from ∼10:51 to ∼10:58 UT. The footpoint of the jet appears at the umbra–penumbra region of the active region (NOAA number: 12665), above a visible light bridge on the sunspot (seen in the 1600 and 1700 Å AIA channels) which has a magnetic configuration of Hale class β. Figure 3 shows the background-subtracted peak intensity of the jet, where the reference time ∼10:45 UT was taken as the background. The jet's spire exhibits a bifurcation as it erupts into two components (see, e.g., Shen et al. 2012). Two artificial slits were used along the path of the two components of the jet's bifurcated spire to assess their propagation (see, e.g., Mulay et al. 2016), as indicated by the red and blue dashed lines in the left panel of Figure 3. The right panel of Figure 3 shows stack plots of distance against time for the erupting jet plasma from the two bifurcated components. At each point in space, the onset of the jet was estimated when the intensity surpassed a threshold of ten times the background level. The jet onset time and the distance along the slit were used to obtain the start time and plane-of-sky speed of each bifurcated component. The first, southern, component (indicated by the blue dashed line in Figure 3) occurs around 10:54:40 UT and has a speed of ∼650 km s$^{-1}$. The second, northern, component (indicated by the red dashed line in Figure 3) starts around 10:56:40 UT, two minutes after the first component, and was found to have a speed of ∼660 km s$^{-1}$. The onset times of the two bifurcated components are strongly correlated to the two peaks shown by the GOES flux density measurements in the right panel of Figure 1.

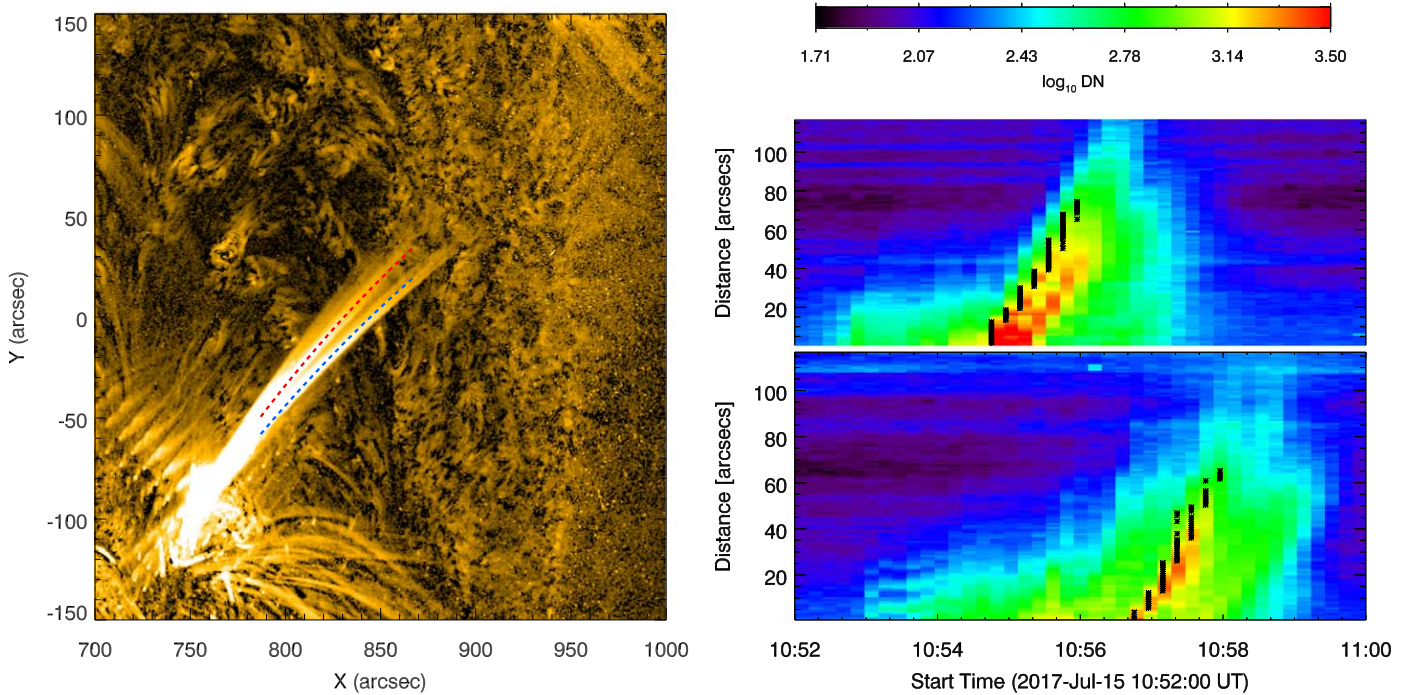

**Figure 3.** The left panel shows the jet (using AIA 171 Å data) with the background subtracted. Two artificial slits (red and blue dashed lines) highlight the two ejections of plasma. The top right panel shows the AIA 171 Å stack plot along the blue artificial slit, whereas the bottom right panel shows the stack plot along the red artificial slit, displaying the propagation of the jet as a function of time. The black stars indicate the times where the intensity surpassed the background level by a factor of 10, used to find the speed of the bifurcated jet components. The first component (blue slit and top right panel) has a speed of ∼650 km s$^{-1}$ while the second component (red slit and bottom right panel) has a speed of ∼660 km s$^{-1}$.

Figure 4 illustrates the spatial and temporal evolution of the two CME fronts with respect to the Sun, as well as an indication of the location of the Type II emissions (green cross). Open magnetic fields that form a thin streamer are visible in the top panels of Figure 4. This streamer is present long before and after the studied activities, and seems to have been formed during an earlier eruption from the active region of interest. The angular width of the CMEs (i.e., their spatial span with respect to the solar center) was measured in the plane of the sky using LASCO/C2 images. The front that appears first in the C2 FOV, as seen in Figure 4, was found to have an average angular width of ∼14°. The front that appears later in the C2 FOV—and north of the earlier front—was found to have an average angular width of ∼5°. As such, the front that appears first will be referred to as the broader front, whereas the front that appears later will be referred to as the narrower front. It should, however, be noted that since both CMEs have an apparent angular width of ≲15°, they classify as narrow CMEs (Gilbert et al. 2001). The average plane-of-sky speeds of the CMEs were estimated by tracking several features across the C2 FOV and applying a linear fit. The broader CME has an average plane-of-sky speed of ∼700 km s$^{-1}$, whereas the narrower CME front has an average plane-of-sky speed of ∼560 km s$^{-1}$. The southern parts of the narrower front and the northern parts of the broader front, however, appear to propagate with an average plane-of-sky speed of ∼470 km s$^{-1}$ in the C2 FOV, likely due to the interaction between the two structures.

It is believed that the two CMEs are the result of a single eruption—the jet eruption—as neither an erupting flux rope or any coronal dimming could be identified within a relatively short time of the jet's eruption. This implies that both CMEs are the product of the observed jet, a behavior similar to the one first reported by Shen et al. (2012) during a blowout-jet eruption. We believe that it is the eruption of the jet's spire into the two bifurcated components that causes the two observed CME fronts, with the first bifurcated component causing the broader CME front and the second bifurcated component causing the narrower CME front which appears at a later time. The two CME fronts, however, show differences in the way they propagate away from the Sun (see Figure 4). The narrower front seems to trace the open magnetic fields that form the streamer, clearly visible in the top panels of Figure 4. The ejection maintains its narrow front and does not appear to disturb the streamer as it traces the path laid out by the open magnetic fields, appearing to be constrained by the streamer. Unlike the narrower front, the broader front appears to deflect away toward the south and is the first to dissipate into the coronal background.

It was noticed that several jet eruptions with similar behavior to the one presented occurred throughout the day. Some examples of these other eruptions include a jet at ∼12:37 UT, at ∼14:43 UT, at ∼16:26 UT, and one at ∼23:09 UT. These jets appear to have the same footpoint location on the edge of the active region as the jet presented in this study, and when ejecta emerged in the C2 FOV, they traced the same streamer structure. The described characteristics—the tracing of a streamer and the repetitive nature of eruptions—resemble that of a "streamer-puff" CME (Bemporad et al. 2005; Panesar et al. 2016; Sterling 2018). First identified as a new variety of CMEs by Bemporad et al. (2005), streamer-puff CMEs are a type of narrow CMEs that move "along the streamer, transiently inflating the streamer but leaving it intact." The Type II emissions appear to have occurred on the northern side of the streamer-puff CME, likely at the flank (see Figure 4).

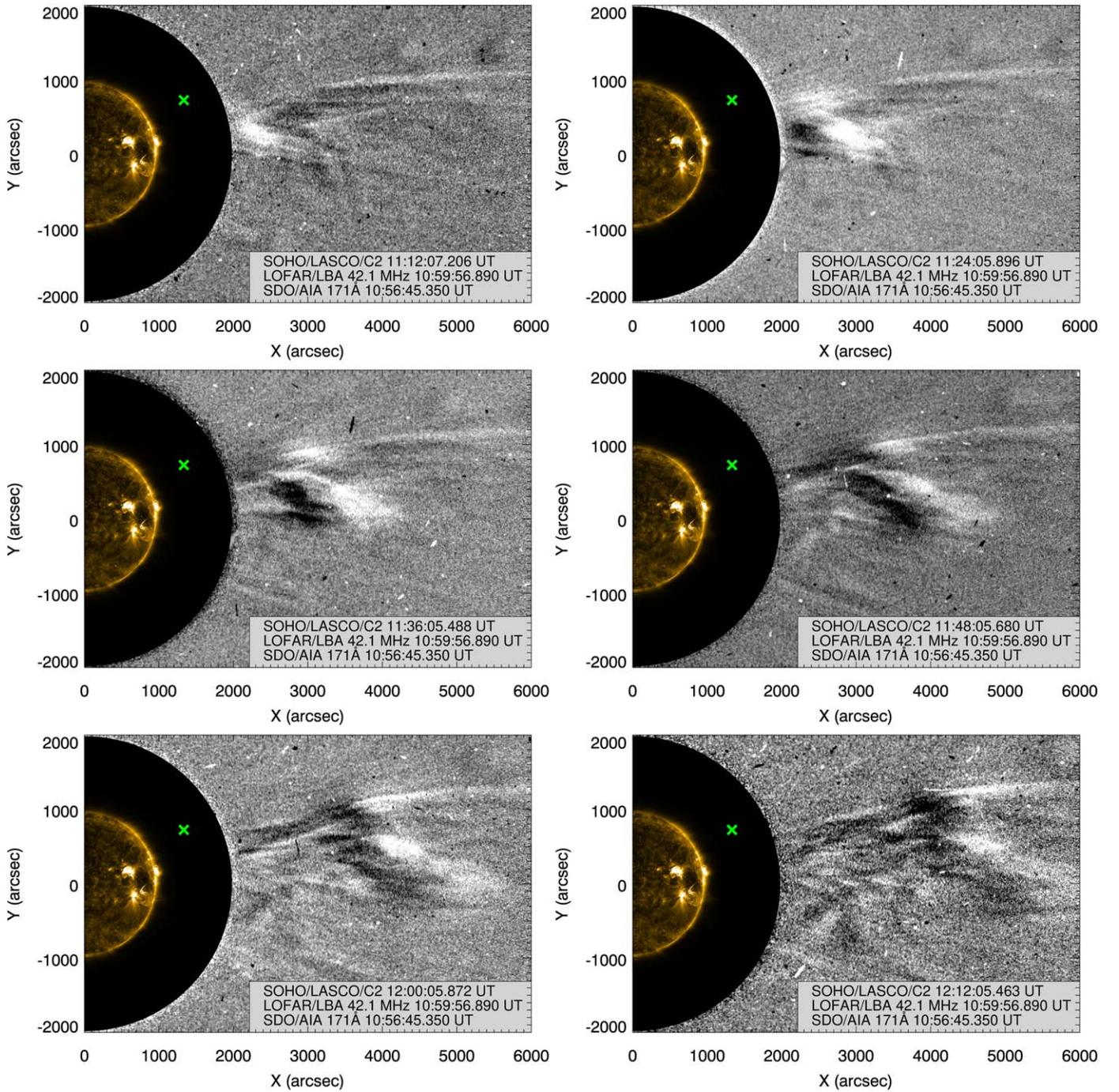

**Figure 4.** Combination of multiwavelength observations from 2017 July 15 that were temporally and spatially related to the Type II radio emissions. The solar surface at the time of the jet's eruption is shown in EUV using SDO/AIA 171 Å data. The green cross illustrates the position of the Type II radio sources as observed by LOFAR, and white-light running-difference images from SOHO/LASCO/C2 are used to depict the motion of CME fronts. The panels depict the consecutive temporal evolution of the CMEs—given the ~12 minutes cadence of LASCO/C2—as they propagate away from the solar surface. Brighter structures in LASCO's FOV reflect relative increases in intensity whereas darker structures reflect relative decreases in intensity. The black disk represents the occulting disk of the C2 coronagraph.

### 3.3. Imaging of the Radio Bursts

Given that the aim of this study is to identify the mechanism resulting in the observed transitioning Type II burst, examining the apparent location of the radio emission sources is of high importance as they can provide an insight into the motion of their exciter. We have therefore employed LOFAR's unprecedented observing capabilities to image the behavior of the emission sources before, during, and after the transition of the

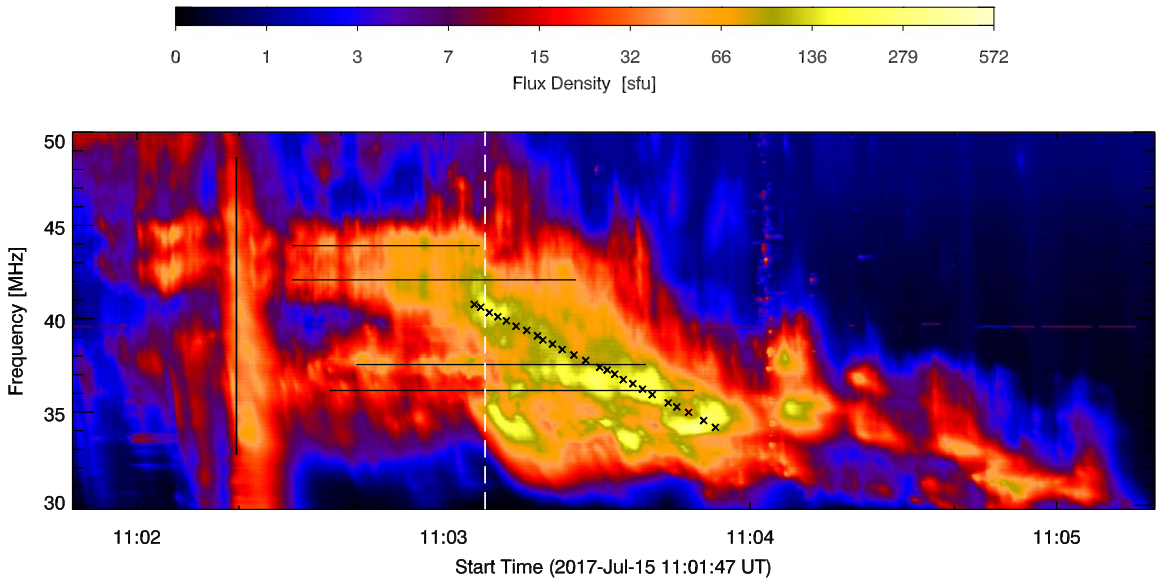

**Figure 5.** Annotated version of the dynamic spectrum shown in Figure 2. The white dashed line indicates the defined time of transition from stationary to drifting Type II emissions, taken to be at 11:03:08.150 UT. The black horizontal lines indicate the single-frequency slices taken for each subband in order to produce emission images before, during, and after the transition between the two states. For the highest-frequency subband, no data was selected past the transition time. The black crosses illustrate the points at which the drifting part of the Type II burst was imaged. The black vertical line at ∼11:02:20 UT indicates the frequencies at which Type III sources were imaged.

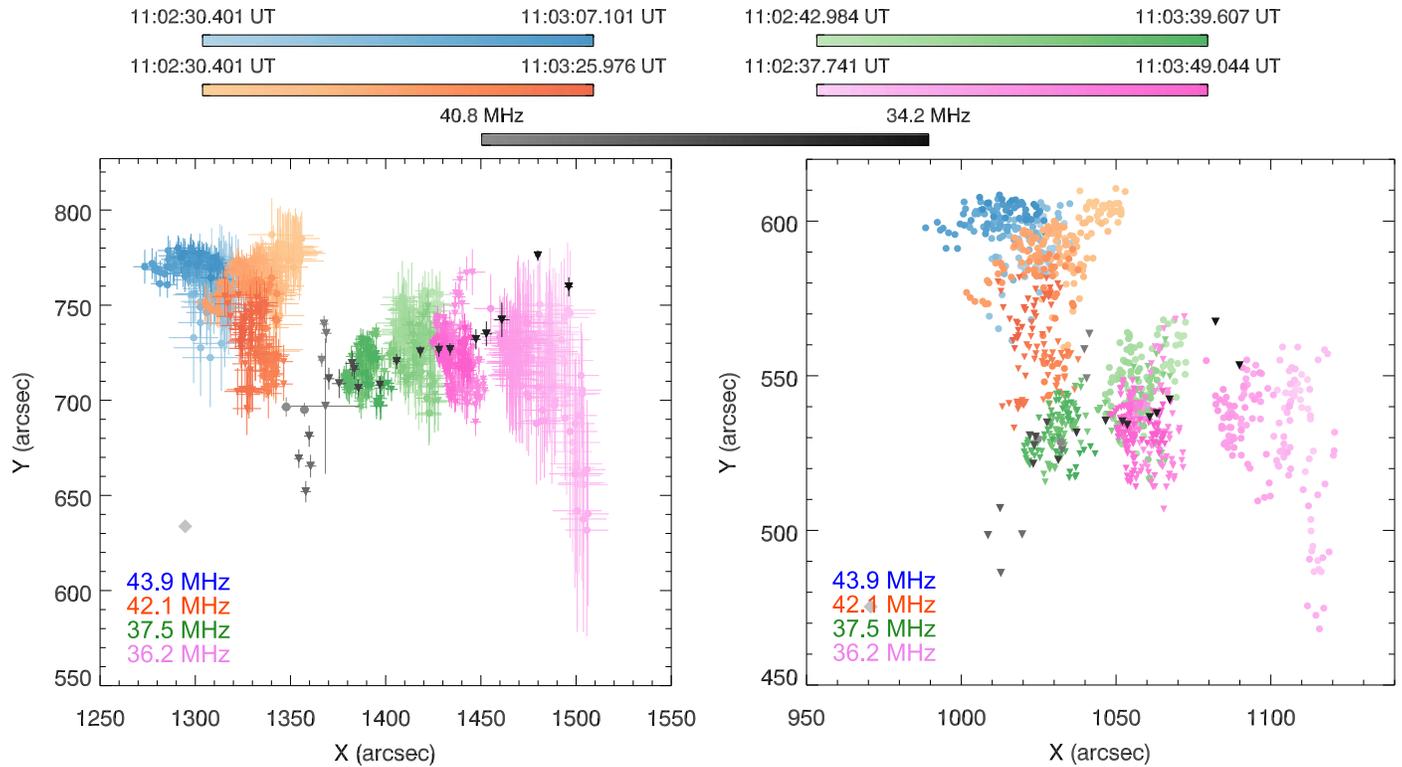

**Figure 6.** Estimated locations of the radio emission sources for each subband, as shown in Figure 5. The left panel displays the centroid locations with their associated errors (one standard deviation) obtained from the 2D elliptical Gaussian fits, whereas the right panel displays the radially corrected (for scattering-induced shift) centroid locations without errors to highlight the motion of the sources. The subband at 43.9 MHz is depicted in a blue color scheme, the subband imaged at 42.1 MHz is depicted in an orange color scheme, the subband imaged at 37.5 MHz is shown in a green color scheme, and the subband imaged at 36.2 MHz is shown in a pink color scheme. The color gradient represents a progression from earlier times (lighter) to later times (darker). Gray centroids illustrate the motion of the drifting Type II emissions, starting from ∼40.8 MHz (light gray) until ∼34.2 MHz (dark gray). Sources represented by a circle occurred before the defined transition time, whereas the ones represented by a downward-facing triangle occurred afterwards. As indicated in Figure 5, no sources past the transition time were imaged for the subband imaged at 43.9 MHz (blue color scheme). Gray diamonds illustrate the central location of the LOFAR beams.

Type II burst from a stationary to a drifting state. Two-dimensional (2D) elliptical Gaussian fits were applied for the determination of the centroid positions—used as a proxy for the location of the radio sources—and the associated one-standard-deviation errors are used as the uncertainty in the estimations (see, e.g., Kontar et al. 2017; Chrysaphi et al.

2018). The sources appear at projected heliocentric distances below ∼1.8 $R_\odot$ which are outside the C2 FOV, thus the CMEs could not be observed at the same heights as the radio emissions (see Figure 4). Nevertheless, the Type II sources appear to have originated close to the top flank of the CME when the CME was closer to the Sun, given that they occur ∼10 minutes before the first CME structures appear in the C2 FOV.

The transition between the stationary and drifting Type II structures occurs around 11:03:08 UT, as can be seen in the dynamic spectrum (Figures 2 and 5). The defined transition time is indicated by the white dashed line in Figure 5. It should, however, be emphasized that a single transition time is merely defined in order to be used as a guiding point in the forthcoming analysis of the emission source locations (see Figure 6); the transition from a stationary to a drifting burst takes place over a few seconds.

In order to examine how—if at all—the motion of the sources changes at the moment of the shock's transition from standing to moving, we image the emissions at multiple moments in time and at a single frequency. Four frequencies are chosen to represent each of the four subbands observed during the stationary Type II part. For the higher-frequency pair of subbands, data was taken at 43.9 and 42.1 MHz for the higher- and lower-frequency components, respectively. Similarly, for the lower-frequency pair of subbands, data was taken at 37.5 and 36.2 MHz. The data at these frequencies was not only selected during the stationary part of the Type II burst, but also at times past the defined transition time so that the behavior of the radio sources before, during, and after the transition could be imaged. The temporal range of these single-frequency slices is indicated by the black horizontal lines in Figure 5. Data points whose flux did not exceed 1% of the maximum flux value of the observation were omitted in order to eliminate the possibility of imaging background noise emissions. For the highest-frequency subband (imaged at 43.9 MHz), none of the drifting Type II emissions beyond the transition point could be confidently related to that subband, thus only data before the transition time was imaged (see Figure 5). By imaging each subband at a single frequency we eliminate the effect of the frequency-dependent radio-wave propagation effects like scattering (see Kontar et al. 2017 and Chrysaphi et al. 2018) on the apparent motion of the sources, meaning that the inferred motion of each subband is purely temporal. In other words, the relative source motions observed within each subband are related to the driver of the radio emissions. However, the absolute location of each of the single-frequency subbands is distorted by radio-wave scattering effects, which cause sources to shift radially away from the solar center, with the shift increasing with decreasing emission frequency (see, e.g., Chrysaphi et al. 2018 for details).

Figure 6 shows the estimated source locations and their apparent temporal evolution for the selected data from each of the four single-frequency slices, as well as the source locations obtained by imaging the drifting part of the Type II burst (see the black crosses in Figure 5). The left panel shows the centroid locations and their associated errors, whereas the right panel shows the corrected for scattering-induced shift centroid locations without error bars (which would be the same as in the left panel) for a clearer illustration of the change in motion. The heliocentric source locations were corrected using the analytical method derived by Chrysaphi et al. (2018) (assuming $\epsilon^2/h = 4.5 \times 10^{-5}$ km$^{-1}$ and fundamental emission), where further details can be found. The corresponding plane-of-sky X–Y locations were obtained through a simple trigonometric relation, given that the angle between the x-axis and the source remains the same during the radial correction for the scattering-induced shift. Each subband is presented in a different color scheme with lighter colors representing earlier times. The subband imaged at 43.9 MHz and before the defined transition time is shown in a blue color scheme. The subband imaged at 42.1 MHz is shown in an orange color scheme, the subband imaged at 37.5 MHz is shown in a green color scheme, and the subband imaged at 36.2 MHz is shown in a pink color scheme. Centroids represented by a circle indicate that the sources occurred before the transition time, whereas the ones represented by a downward-facing triangle occurred after the transition time (see the white dashed line in Figure 5). The sources from the drifting part of the Type II burst are illustrated with a gray color scheme which represents frequencies from ∼40.8 MHz (light gray) to ∼34.2 MHz (dark gray). It should be emphasized that the drifting Type II emission sources appear to propagate away from the Sun, as expected.

Upon comparison of the centroids indicated with circles and those indicated with downward-facing triangles, it can be seen from Figure 6 that a jump in the collective position of the sources of each subband occurs around the defined transition time. Furthermore, the sources of each subband do not appear to be gathered in a single location as expected for emissions excited by standing shocks, demonstrating that the structure exciting the Type II emissions may not be completely stationary. The color progression implies that the sources move toward the solar surface as time passes. The fact that a single frequency was used to image each subband but a spatial evolution is observed can be interpreted as the apparent motion toward the Sun of a structure with a constant density-to-background-density ratio, thus affecting the imaged position of the sources but not the emitting frequency. Due to projection effects, a source moving away from the solar center at an angle to the observer's line of sight (LOS) can sometimes appear as if it is moving toward the solar center in the plane of the sky. See the Appendix for details on when this becomes the case.

The corrected source locations of the drifting Type II emissions were used to estimate the plane-of-sky speed of the shock wave, found to be ∼840 km s$^{-1}$. It should be noted that if the correction for scattering-induced radial shifts is omitted, the resulting shock speed is unreasonably high (∼2220 km s$^{-1}$) given the estimated CME speed. The correction for scattering effects reduces the estimated speed as the heliocentric distances of lower-frequency sources are decreased more than those of higher-frequency sources. This in turn decreases the collective spatial expansion of the sources over the given time. For comparison, we also estimated the shock speed using a coronal density model and the estimated drift rate of the burst (∼−0.14 MHz s$^{-1}$) obtained from the dynamic spectrum (see Equation (3) in Chrysaphi et al. 2018). We took the Newkirk (1961) density model that best matched the corrected radial locations (see Figure 5 in Chrysaphi et al. 2018 for comparison) and obtained a speed of ∼760 km s$^{-1}$, which is also reasonable and similar to that obtained using the corrected imaged locations.

Additionally, an examination of the fine structures of the Type II burst revealed intriguing negative as well as positive frequency-drift structures within the stationary Type II part, shown in Figure 7. These fine structures do not resemble the well-known fine structures referred to as "herringbones" that are often observed in Type II bursts (McLean & Labrum 1985). The white-line annotations in Figure 7 highlight the observed fine structures and their altering frequency-drift rates. From left to right, the drift rates of these fine structures were approximated to be −0.25, −0.51, +0.20, −0.93, and +0.41 MHz s$^{-1}$. These emissions of altering frequency drifts indicate the existence of a pulsating driver. Such behavior is reminiscent of the Type II burst reported by Mel'nik et al. (2004) which on average showed no drift but had a "waving backbone." However, the waving backbone reported by Mel'nik et al. (2004) is comprised of structures with different drift rates that individually last over several minutes, whereas in our case, each fine structure only lasts a few seconds. It is possible that these fine structures are related to the irregular surface (or surface fluctuations) of shocks that were recently observed in the interplanetary space (Kajdič et al. 2019) and are known to be important for electron acceleration (Trotta & Burgess 2019).

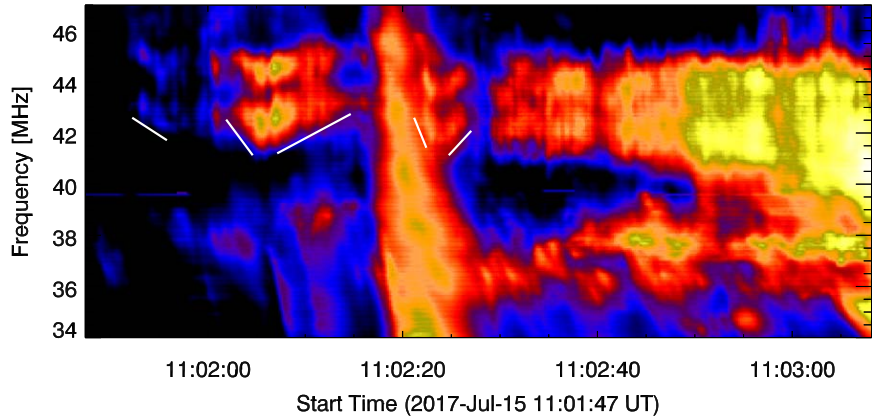

**Figure 7.** A section of the dynamic spectrum shown in Figures 2 and 5, plotted with a different dynamic range to highlight the fine structures within the stationary part of the Type II burst. The five white-line annotations emphasize the altering frequency-drift rates of some of the fine structures that are easily distinguishable. As evident, some of the fine structures have positive drift rates and some negative.

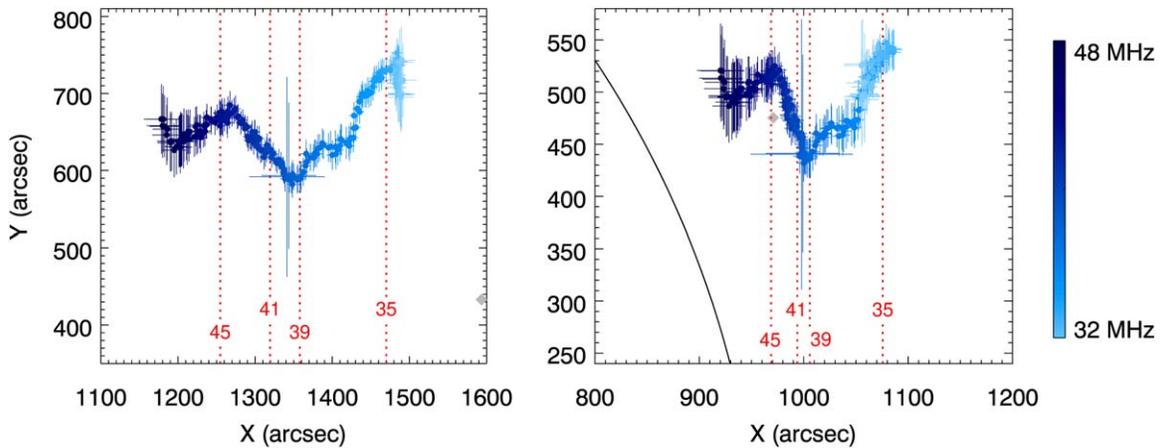

**Figure 8.** Estimated locations of the Type III sources with their associated errors imaged at a single time (∼11:02:20 UT) but multiple frequencies (∼32–48 MHz), as indicated by the black vertical line in Figure 5. The red lines indicate the locations at which the four frequencies representing the bandwidths of the two pairs of Type II burst subbands are emitted (35–39 MHz and 41–45 MHz, see Section 3.1). The left panel shows the apparent centroids locations, whereas the right panel shows the corrected locations for the scattering-induced radial shift. Gray diamonds illustrate the central location of the LOFAR beams, while the solid black curve in the right panel represents the solar limb.

A Type III burst was observed during the stationary Type II emissions. Type III bursts are attributed to electron beams that trace open magnetic fields (Melrose 1990; Reid & Ratcliffe 2014), and their sources tend to follow a smooth curve with higher-frequency sources found closer to the Sun than lower-frequency sources (see, e.g., Zhang et al. 2019). The sources of the observed Type III burst were imaged at a single moment in time, indicated by the black vertical line in Figure 5. An unusual pattern is observed as there are two abrupt shifts in the direction of motion of the sources, rather than a smooth progression. Figure 8 depicts the sources of the Type III burst and the striking shifts in the direction of their motion, with the left panel showing the apparent source locations and the right panel showing the locations corrected for the radial shift caused by scattering. The first shift—where the direction of motion changes toward the south of the Sun—occurs around 45 MHz, and the second shift—where the motion changes back to a northerly direction—occurs around 39 MHz (see the red line annotations in Figure 6). These frequencies (39–45 MHz) coincide with the bandwidth of the three higher-frequency subbands of the stationary part of the Type II burst (see Figure 5).

## 4. Summary and Conclusions

A Type II solar radio burst that experiences a transition between a standing and a drifting state has been reported for the first time. Band splitting was observed during the stationary Type II emissions, as well as fine structures of both negative and positive frequency drifts. A Type III burst with an unusual source propagation away from the Sun was also observed during the emissions forming the stationary Type II part. This study focused on gathering observational evidence in order to understand the mechanism that generated radio emissions manifesting into the morphological structure described as a transitioning Type II burst.

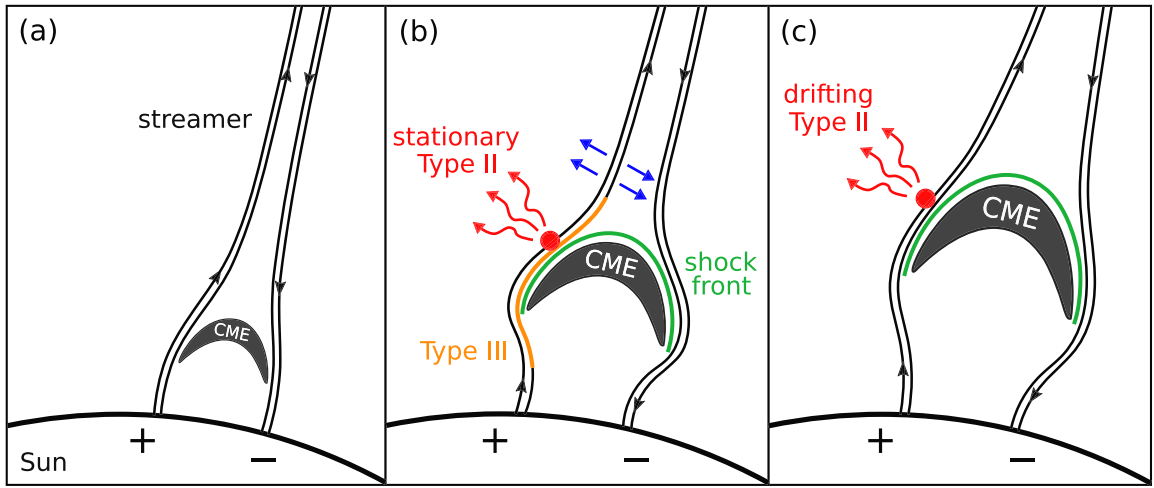

**Figure 9.** Schematic illustration of the key phases of the mechanism generating the observed radio emissions. Panel (a) illustrates the streamer-puff CME that was formed following the jet's eruption. Panel (b) illustrates the CME as it propagates along the streamer and expands, as well as the shock front forming ahead of it (green curve). The streamer undergoes an abrupt local expansion and the consequent compression by the shock results in the stationary Type II emissions (shown in red), as regions of the shock front are halted by the streamer. The interplay between the streamer and the CME causes the streamer to pulsate (blue arrows) which is reflected in the negative and positive frequency-drift fine structures observed during the stationary Type II emissions (see Figure 7). An electron beam traces the curved magnetic fields confining the streamer and results in Type III burst emissions (orange curve). Panel (c) shows the moment that the steamer succumbs to the CME's expansion and allows it to smoothly propagate away from the Sun, while the compression between the streamer and the moving shock excite the drifting Type II emissions (shown in red).

High-resolution imaging spectroscopy of the radio emissions was obtained with LOFAR. A coronal jet was observed in AIA EUV and GOES X-ray data in close temporal and spatial proximity to the Type II emissions. It was found that the spire of the jet experiences bifurcation which is believed to have resulted in two CME eruptions observed in the LASCO/C2 FOV, with one being narrower than the other. The two CMEs showed a difference in behavior and evolution, as detailed in Section 3.2. The apparent Type II sources appeared to be located near the northern flank of the narrow CME, away from any potential interaction between the edges of the narrower CME and the southern, broader, CME. The narrow CME traced a thin streamer as it propagated into the corona but, crucially, did not disturb the streamer during its passage. Multiple jet eruptions originating from the same part on the edge of the active region as the jet presented in this study were continuously observed on 2017 July 15, as mentioned in Section 3.2. The ones that manifested into coronal ejecta in the C2 FOV also followed the same streamer traced by the studied eruption. Due to these characteristics, the narrow CME front was identified as a streamer-puff CME (Bemporad et al. 2005; Sterling 2018).

The observations suggest that the streamer-puff CME is the driver of the shock wave exciting the transitioning Type II emissions. In Figure 9, we present a schematic illustration of the generation mechanism and its key phases which we believe resulted in the observed radio emissions (see Figure 2). Due to the jet's eruption and the presence of the streamer, a streamer-puff CME is formed, indicated in panel (a). Once the CME speed exceeds the local Alfvén speed (which decreases with heliocentric distance) a shock front is formed ahead of the CME (McLean & Labrum 1985), indicated by the green curve in panel (b). The shock wave presses against the open magnetic fields forming the streamer, and the streamer undergoes a localized expansion around the flanks of the CME, but not yet near the nose of the CME. Regions of the shock are halted by the interplay with the streamer and effectively behave as a standing shock structure. We believe that it is at this stage (see panel (b)) that three different but nearly simultaneous actions take place:

1. The compression occurring during the interaction between the shock front and the streamer excites radio emissions (shown in red). This is the moment in time at which the stationary Type II burst is formed, i.e., when the CME causes the streamer to quickly locally expand, but before the undisturbed parts of streamer (upstream of the shock) expand enough to allow for the smoother transition of the CME.
2. During this interaction, the negative and positive frequency-drift fine structures within the stationary Type II part are formed (see Figure 7). They are interpreted as the result of the streamer pulsating (blue arrows) as it interplays with the expanded shock.
3. An electron beam traces the open magnetic fields confining the locally expanded streamer, forming a Type III burst (orange curve) of which the source locations reflect the curving exhibited by the magnetic fields (see Figure 8).

Figure 9(c) depicts the final stage where the CME forces the streamer to succumb to its expansion, even around the nose of the CME. We believe that at the moment during which the streamer inflates, the region of the shock front exciting radio emissions transitions from a standing shock to a drifting shock, and the streamer structure that was pulsating abruptly "jumps" to a new location, allowing the CME to smoothly travel along the streamer. In other words, this is the moment at which the transition occurs and the jump in the Type II sources is observed (see Figure 6), making the sources appear as though they are propagating toward the Sun when in fact they could be moving away from the Sun in the *z*-direction (see the Appendix). Finally, as the shock wave smoothly moves away from the Sun and continues to expand, the constant compression with the streamer excites the drifting Type II emissions (shown in red).

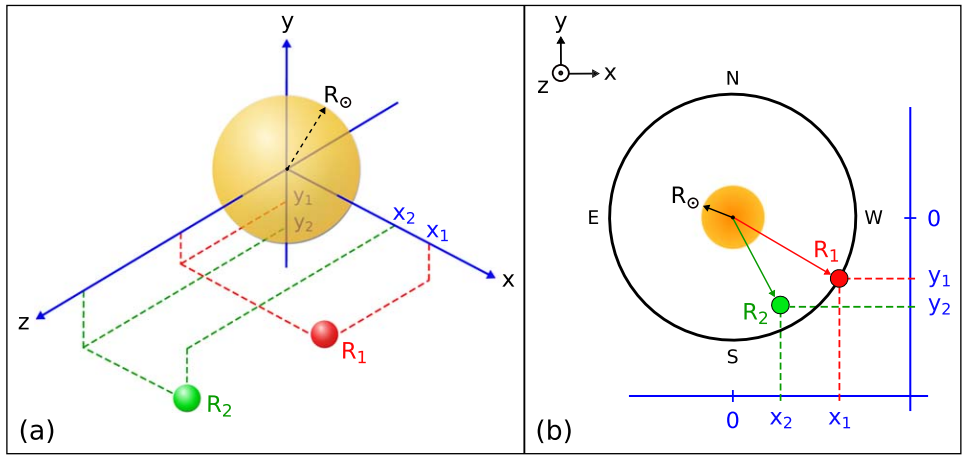

**Figure A1.** Illustration of projection effects on 2D plane-of-sky observations, where the observer's LOS is parallel to the *z*-axis. Panel (a) is a 3D depiction of two sources emitted at $\boldsymbol{R}_1$ (red source) and $\boldsymbol{R}_2$ (green source), where $|\boldsymbol{R}_2| > |\boldsymbol{R}_1|$. Panel (b) depicts the plane-of-sky projection (*xy*-plane), where the green source is observed closer to the solar center than the red source, whereas in reality it is farther away.

In conclusion, we have reported a new subclass of Type II bursts, the "transitioning" Type II bursts. Using a combination of multiwavelength observations, we have presented a scenario explaining the observed radio emissions. A jet eruption caused a steamer-puff CME which produced both the stationary and drifting Type II structures as it interacted with the streamer.

N.C. was supported by the STFC grant ST/N504075/1. H. A.S.R. and E.P.K. were supported by the STFC grant ST/P000533/1. The authors acknowledge the support by the international team grant (http://www.issibern.ch/teams/lofar/) from ISSI Bern, Switzerland. This paper is based (in part) on data obtained from facilities of the International LOFAR Telescope (ILT) under project code LC8_027. LOFAR (van Haarlem et al. 2013) is the Low-Frequency Array designed and constructed by ASTRON. It has observing, data processing, and data storage facilities in several countries, that are owned by various parties (each with their own funding sources), and that are collectively operated by the ILT Foundation under a joint scientific policy. The ILT resources have benefited from the following recent major funding sources: CNRS-INSU, Observatoire de Paris and Université d'Orléans, France; BMBF, MIWF-NRW, MPG, Germany; Science Foundation Ireland (SFI), Department of Business, Enterprise and Innovation (DBEI), Ireland; NWO, The Netherlands; The Science and Technology Facilities Council, UK; Ministry of Science and Higher Education, Poland. The SOHO/LASCO data used here are produced by a consortium of the Naval Research Laboratory (USA), Max-Planck-Institut fuer Aeronomie (Germany), Laboratoire d'Astronomie (France), and the University of Birmingham (UK). SOHO is a project of international cooperation between ESA and NASA. The authors would also like to thank the SDO/AIA team and the GOES/XRS team for the data.

## Appendix
## Projection Effects

Sources that propagate away from the solar center and at an angle to the observer's LOS, can appear in the plane of the sky as if they are propagating toward the solar center due to projection effects. This is depicted in Figure A1, where the LOS of the observer is along the *z*-axis, thus the *xy*-plane is the plane of the sky. Panel (a) is a three-dimensional (3D) illustration of two sources emitted at heliocentric distances $\boldsymbol{R}_1 = (x_1, y_1, z_1)$ (red source) and $\boldsymbol{R}_2 = (x_2, y_2, z_2)$ (green source), where $|\boldsymbol{R}_2| > |\boldsymbol{R}_1|$. However, in the plane-of-sky projection, the green source is observed closer to the solar center than the red source, as seen in panel (b), when $R_2^2 > R_1^2$ and $x_2^2 + y_2^2 < x_1^2 + y_1^2$.

**ORCID iDs**

Nicolina Chrysaphi https://orcid.org/0000-0002-4389-5540
Hamish A. S. Reid https://orcid.org/0000-0002-6287-3494
Eduard P. Kontar https://orcid.org/0000-0002-8078-0902